\documentclass[aps,prb,twocolumn,amsmath,amssymb,floatfix]{revtex4-1}
\usepackage{tabularx}
\usepackage{bm}
\usepackage{euscript}
\usepackage{graphicx}
\usepackage{color}
\usepackage{amsfonts}
\usepackage{exscale}
\usepackage{amsbsy}
\usepackage{subfigure}
\usepackage{textcomp}

\begin{document}
 
\title{Charge and spin collective modes in a quasi-1D model of Sr$_2$RuO$_4$}
\author{Suk Bum Chung$^1$, S. Raghu$^{1,2}$, Aharon Kapitulnik$^1$, and Steven A. Kivelson$^1$}
\affiliation{$^1$Department of Physics, Stanford University, Stanford, California 94305, USA}
\affiliation{$^2$Stanford Institute for Materials and Energy Science, SLAC National Accelerator Laboratory, Menlo Park, California, 94025, USA}
 
\date{\today}
 
\begin{abstract}
Given that Sr$_2$RuO$_4$ is a two-component p-wave superconductor, there exists the possibility of well defined collective modes corresponding to fluctuations of the relative phase and spin-orientation of the two components of the order parameter. We demonstrate that at temperatures much below $T_c$, these modes have energies small compared to the pairing gap scale if the superconductivity arises primarily from the quasi 1D ($d_{xz}$ and $d_{yz}$) bands, while it is known that their energies become comparable to the pairing gap scale if there is a substantial involvement of the quasi 2D ($d_{xy}$) band. Therefore, the orbital origin of the superconductivity can be determined by measuring the energies of these collective modes.
\end{abstract}
 
\maketitle
 
\section{Introduction}
The layered perovskite material Sr$_2$RuO$_4$ has attracted strong interest for over a decade because of experiments indicating that it is a spin-triplet superconductor \cite{ISHIDA1998, DUFFY2000, NELSON2004, KIDWINGIRA2006} with spontaneously broken time-reversal symmetry \cite{LUKE1998, XIA2006}. The only state that possesses both of these properties in a tetragonal system with spin-orbit coupling 
is the chiral $p$-wave state, the electronic analogue of $^3$He-A \cite{ANDERSON1961, RICE1995}. In a single-band, quasi-two dimensional system, this state is expected to be a topological superconductor: it is fully gapped and has topologically protected Majorana fermion zero modes in vortex cores and along its edge.

However, the observation of power laws in specific heat \cite{NISHIZAKI1999} and NMR \cite{ISHIDA2000}, the absence of electric currents along edges and domain walls \cite{Bjornsson2005}, and the absence of a split transition in the presence of an in-plane magnetic field \cite{MAO2000} are sharply inconsistent with the theoretically expected properties of a simple chiral superconductor. Motivated by these inconsistencies, three of us \cite{RAGHU2010} proposed that the multi-band nature of the material is essential.

\begin{figure}
\includegraphics[width=3.0in]{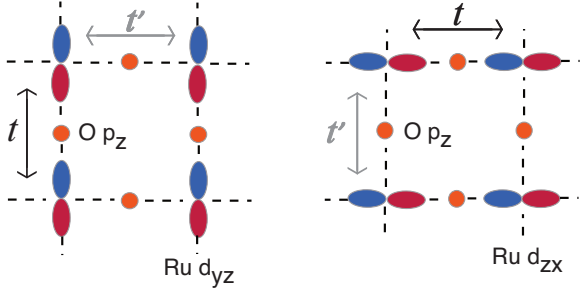}
\caption{Schematic view of the Ru $d_{xz}$ and $d_{yz}$ orbitals in the Ru-O plane. The wave-function overlap is larger along the direction of the black arrows than along the direction of the gray arrows, giving us $t \gg t'$, and therefore, the quasi-1D nature of the bands originating from these orbitals.}
\label{FIG:1d}
\end{figure}

The electronic structure of this system is derived from the Ru t$_{2g}$ electrons ($d_{xz}, d_{yz}$, and $d_{xy}$).  These orbitals produce three Fermi surfaces, denoted $\alpha$,  $\beta$, and $\gamma$.  The $\alpha$ and $\beta$ surfaces are derived from the $\{  d_{xz}, d_{yz} \}$ orbitals and are quasi-one dimensional (1D), whereas the $\gamma$ surface is derived primarily from the $d_{xy}$ orbital and is quasi-two dimensional (2D) (see Fig. \ref{FIG:1d}).    
Because of the strong differences in character between the different bands, superconductivity in this system 
is likely derived primarily from either the $d_{xy}$, or $\{ d_{xz}, d_{yz} \}$ orbitals: in either scenario the ``active" electrons induce superconductivity in the remaining ``passive" subset via the proximity effect \cite{AGTERBERG1997}.
However, to the extent that the proximity effect is weak, there is a range of circumstances in which the experimental observations reflect superconducting properties of mainly the active orbitals.
 
There are sharp distinctions between the two possibilities for the active electrons. For example, the $p_x + i p_y$ state obtained {\it only} from the $\alpha, \beta$ bands is topologically trivial because the $\alpha$ and $\beta$ bands form, respectively, a hole and electron pocket leading to a net zero Chern number\cite{RAGHU2010}. If these were the active electrons, the system behaves as a topologically trivial system near $T_c$. At $T=0$, superconductivity occurs on all three bands and is therefore topologically non-trivial; however, the size of the induced gap on the passive $\gamma$ Fermi surface can be substantially smaller than those on the $\alpha, \beta$ surfaces, making it difficult to experimentally detect the topologically non-trivial character of the ground state. This scenario could explain, for instance, the absence of any detectable edge currents in the system\cite{Bjornsson2005}. By contrast, when $\gamma$ is the active band, the system behaves as a topological superconductor even near $T_c$. Recently, it was shown that there is an intrinsic Kerr response near T$_c$ only when $\{ \alpha, \beta \}$ are the active bands\cite{Taylor2012, Wysoki2012} . In this paper, 
as part of a further exploration of the experimentally accessible properties that could distinguish between the two cases,
we study the qualitative differences in the character of the
charge and spin collective modes.

 
 
 
In forming the $\alpha$ and $\beta$ bands, the $d_{xz}$ and $ d_{yz}$ orbitals are coupled
to each other only via spin-orbit coupling and second-neighbor hopping terms, 
both of which are relatively weak\cite{Haverkort2008} in Sr$_2$RuO$_4$. 
The limit in which these orbital mixing terms vanish defines a {\it multicritical} point at which 
the superconducting state breaks a higher $\left[SO(3)_{\rm spin} \times U(1)_{\rm charge} \right]^2$ symmetry. 
Proximate to this multicritical point, {\it i.e.} for weak orbital mixing, there are low energy ``almost Goldstone modes''
associated with fluctuations of the relative phase and spin orientation of the $x$ and $y$ components of the superconducting order parameter. By contrast, the same collective modes have\cite{Higashitani2000, Kee2000} energies of order $\Delta_0$ when $\gamma$ is the active band.
 
This paper is organized as follows. In section II, we discuss the general physics of the collective modes in a multi-component superconductor at a qualitative level and describe the general form of the non-linear sigma model (NLSM) of the system. In sections III-V, we derive the NLSM from the microscopic physics of the quasi-1D model and obtain the gaps of the collective modes. Lastly, we discuss schemes for detecting the modes and consider their broader implication for the superconducting properties of the system.

\section{
Collective modes in multi-
component superconductors}
\subsection{Qualitative discussion} A superconductor described by a multi-component order parameter ($p_x$ and $p_y$ in the present context), will have collective mode excitations associated with the relative phase difference, $\phi_-\equiv\theta_{x}-\theta_{y}$.
At zero temperature, such a mode would be expected to have a frequency $\hbar\omega_0\sim \sqrt{{\cal J} /\chi}$, where $\chi \sim N(0)$ is the compressibility ($N(0)$ is the density of states at the Fermi energy) and ${\cal J}$ is the second derivative of the condensation energy with respect to $\phi_-$. Given that the condensation energy $\sim N(0) |\Delta_0|^2$ where $|\Delta_0|$ is the root mean squared gap magnitude, this generally means that $\hbar\omega_0 \sim |\Delta_0|$. Similar considerations apply to fluctuations in the relative orientation of the spins ({\it i.e.} the $d$-vector) in a two-component triplet superconductor.
  
By contrast, 
if the two components of the order parameter 
are associated with different orbitals, {\it i.e.} the $p_x$ component with the $d_{xz}$ and the $p_y$ with $d_{yz}$ orbital respectively, and if mixing between the different orbitals were absent, then collective fluctuations of the relative phase and spin-orientation would be gapless.
The orbital-mixing terms, which we schematically denote as $\delta H$ (and which we discuss in detail below), result in a non-vanishing dependence of the condensation energy on $\phi_-$.
As a result, if the superconductivity arises primarily from the quasi-1D bands, then the relative phase mode is expected to have an energy $\hbar\omega_0 \approx \gamma \Delta_0$, where $\gamma$ vanishes continuously as $\delta H$ tends to zero. 
Naturally, similar considerations apply to the relative spin orientational fluctuations. As we explain below, since appreciable orbital mixing occurs only where the bands cross one another, $\gamma \ll 1$ even when the characteristic scale associated with $\delta H \gg \vert \Delta_0 \vert$.
 
 
\subsection{Non-linear sigma model}
To analyze the
low-lying collective modes in a two-band spin-triplet superconductor 
we consider the non-linear sigma model (NLSM) valid deep inside the superconducting phase. We 
express the order parameter as
\begin{equation}
\label{delta}
\Delta_{\alpha ; ss'}(\bm k) = \Delta f_\alpha ({\bf k}) e^{i\theta_\alpha} (i\sigma_2 {\bf \hat{d}}_a \cdot {\bm \sigma})_{ss'},
\end{equation}
where 
$\alpha = x$, $y$ 
labels the two components of the order parameter,
$s,s'$ the spin indices of the two electrons forming the Cooper pair, $\Delta$ the order parameter amplitude, $f_\alpha({\bf k})$ the 
pair wave-function which is determined by the microscopic form of the pairing interaction, and ${\bf \hat{d}}_\alpha$ a real unit vector in spin-space. 
In an $N$-orbital basis, $f_\alpha(\bm k)$ (and consequently $\Delta_{\alpha,ss'}$) are $N\times N$ matrices which transform as the $x$ and $y$ components of a vector of the tetragonal point group. 
 
The NLSM action is obtained by holding $\Delta$ and $f_{\alpha}(\bm k)$ fixed, and focusing on the 
long-wave-length fluctuations of $\theta_\alpha$ and ${\bf \hat{d}}_\alpha$:
\begin{align}
\mathcal{L} =& \frac{1}{2} \sum_\alpha\left[\mathcal{W}
|\partial_t \theta_\alpha|^2
+
\mathcal{M}
(\partial_t {\bf \hat{d}}_\alpha)\cdot(\partial_t {\bf \hat{d}}
_{\alpha}) +\ldots \right]\nonumber \\
-& \mathcal{J}[\theta_{x} - \theta_{y}, {\bf \hat{d}}_{x} \cdot {\bf \hat{d}}_{y}] +\ldots \nonumber \\
- &\sum_\alpha\Gamma_0
({\bf \hat{d}}_\alpha \cdot {\bf \hat{d}}_{\alpha}/3 - \hat{d}_\alpha^z \hat{d}_\alpha^z) + \ldots
\label{EQ:nlsm}
\end{align}
Here, the first line contains the terms which respect the 
symmetry 
 
of the multicritical system.  The $``\ldots"$ includes terms proportional to spatial-derivatives and higher powers of time derivatives.  Our interest is in the long wavelength limit and we shall neglect spatial derivatves; however, these terms are the only ones which reflect the underlying tetragonal lattice symmetry.    

The second-line includes terms derived from inter-orbital mixing without spin-orbit coupling, and the third line includes spin-orbit coupling terms.  Again, derivative terms are neglected.     Here,  $\mathcal{J}[\phi,x]=\mathcal{J}[\phi+\pi,x]=\mathcal{J}[-\phi,x]=\mathcal{J}[\phi,-x]$.  
Note that, 
there is no 
Goldstone mode resulting from fluctuations of the overall phase $\phi_+\equiv\theta_{xz} + \theta_{yz}$ 
due to the Anderson-Higgs mechanism.
 
In the next section, we define a microscopic model of the superconductivity on the quasi-1D bands and from it derive estimates of the various couplings that appear in Eq. \ref{EQ:nlsm}: Estimates for $\mathcal {W}$ and $\mathcal{M}$, both of which are of order $1/N(0)$, are discussed below Eq. \ref{time}. To first order in the (spin rotationally invariant) orbital mixing terms, the form $\mathcal{J}$ is found in Eqs. \ref{EQ:minApprox} and \ref{EQ:spinJ} to be
\begin{equation}
\mathcal{J}[\phi_-,\hat d\cdot\hat d^\prime] = {\cal J}_0\cos(2\phi_-)\big[ 2(\hat d\cdot\hat d^\prime)^2-1\big]
\label{calJ}
\end{equation}
and an estimate of $\mathcal{J}_0$ is presented in Eq. \ref{EQ:interApprox}. Finally, an expression for $\Gamma_0$ to second order in the spin-orbit coupling is given in Eq. \ref{EQ:dVecPin}.

Because ${\cal J}_0 >0$, the orbital mixing term is minimized either when $\hat d_{x}$ is parallel to $\hat d_{y}$ and $\phi_-=\pi/2$, which corresponds to the chiral superconducting state analogous to the A-phase of $^3$He, or when $\hat d_{x}$ is perpendicular to $\hat d_{y}$ and $\phi=0$ or $\pi$, which corresponds to the time-reversal invariant superconducting analogue of the B-phase of $^3$He.
The degeneracy between these two phases is exact (at mean-field level) in the absence of spin-orbit coupling; this observation and the various periodicity conditions impose general constraints on the form of $\mathcal{J}$, which are satisfied by our present result. The degeneracy between the A and B phases is lifted by spin-orbit coupling: for $\Gamma_0 > 0 (\Gamma_0 < 0)$, the A(B) phases have larger condensation energy.
 
We now proceed to show explicitly how these qualitative considerations apply in a simple, but physically motivated microscopic model of Sr$_2$RuO$_4$.
 
  
\section{Quasi-1D model of {\protect Sr$_2$RuO$_4$} superconductivity}
As a microscopic representation of the problem of the pairing in the quasi-1D bands, we consider an idealized form of the Bougoliubov-de Gennes Hamiltonian for the quasi-particles in a p-wave superconducting state,
\begin{eqnarray}
H_{BdG} = 
H_{multi} + \delta H.
\label{H}
\end{eqnarray}
Here $H_{multi}$ represents the multicritical point model, in which there is no orbital mixing or spin-orbit coupling, and $\delta H$ (assumed small) represents terms which break the higher symmetry of the multicritcal point.
 
\subsection{The multicritical point Hamiltonian}

$H_{multi}$ is the mean-field Bogoliubov-de Gennes Hamiltonian for decoupled $xz$ and $ yz$ orbitals:
\begin{align}
\label{EQ:1D0}
H_{multi} &= \sum_{a=xz,yz}\sum_{{\bf k}s} 
\xi_a ({\bf k}) c^\dagger_{a{\bf k}s} c_{a{\bf k}s}\\
+ \frac{1}{2}&\sum_{a=xz,yz}\sum_{\alpha=x,y}\sum_{{\bf k};ss'}[\Delta^{(a)}_{\alpha; ss'}(\bm k) c^\dagger_{a,{\bf k},s}c^\dagger_{a,-{\bf k},s'}+{\rm h.c.}],
\nonumber
\end{align}
where $c_{a{\bf k}s}$ is the annihilation operator for an electron with the momentum ${\bf k}$, 
orbital index $a = xz,yz$ and 
spin polarization $s$, $\xi_a({\bf k}) = \epsilon_{a{\bf k}} - \mu$, the chemical potential $\mu$ is
set so that these bands are two-third filled \cite{MACKENZIE2003, Haverkort2008}, and $\Delta^{(a)}_{\alpha; ss'}$ is the appropriate orbital diagonal matrix element of the pair-field defined in Eq. \ref{delta}.
We further simplify the model by taking the band-structure in the absence of orbital mixing to be strictly one dimensional,
$\epsilon_{a{\bf k}} = -2t \cos k_a $ 
and the $x$ and $y$ components of the order parameter to originate entirely on the corresponding 1D band,
$\Delta^{(xz)}_{y,ss'} = \Delta^{(yz)}_{x,ss'} =0$,
$\Delta^{(xz)}_{x; ss'}(\bm k) = \Delta \sin(k_x) e^{\theta_x}[i\sigma_2\hat d_x\cdot \vec \sigma]_{ss'}$ and
$\Delta^{(yz)}_{y; ss'}(\bm k) = \Delta \sin(k_y) e^{\theta_y}[i\sigma_2\hat d_y\cdot \vec \sigma]_{ss'}$. While this simplified band and gap structure simplifies the explicit calculations, the qualitative results we have obtained are not affected by the inclusion of moderate transverse (but still orbital-diagonal) components of the hopping matrix. 
 
Because the two bands are related by rotation by $\pi/2$, $H_{multi}$ respects the tetragonal symmetry of the material even though each band is one dimensional.
From the fact that $\Delta_{x}$ lives entirely on the $d_{xz}$ band and $\Delta_y$ 
on the $d_{yz}$ band, 
it further follows that $H_{multi}$ respects a full $\left[SO(3)_{\rm spin} \times U(1)_{\rm charge} \right]^2$ symmetry, so 
there is no 
dependence of the free energy on 
$\theta_\alpha$ or $\hat{d}_{\alpha}$. 
In particular, note that the quasiparticle spectrum is fully gapped not only in the A-phase ({\it e.g.} for $\hat d_x\cdot\hat d_y=1$ and $\theta_x-\theta_y=\pi/2$) and B-phase ({\it e.g.} for $\hat d_x\cdot\hat d_y=0$ and $\theta_x-\theta_y=0$), but also for a $p_{x \pm y}$ ({\it e.g.} $\hat d_x\cdot\hat d_y=1$ and $\theta_x-\theta_y=0,\pi$) which might otherwise have been expected to posses gap nodes.
 
\subsection{Orbital mixing and spin-orbit coupling}
$\delta H$ contains all permissible terms in the quasi-particle Hamiltonian which break the $\left[SO(3)_{\rm spin} \times U(1)_{\rm charge} \right]^2$ symmetry of the multicritical point, of which the most important are band-structure terms that mix the two orbitals,
\begin{align}
\label{EQ:orbHyb}
\delta H &= \sum_{{\bf k}s} \lambda_{\bf k}(c^\dagger_{xz,{\bf k},s} c_{yz,{\bf k},s}+{\rm h.c.})\\
&+ \eta\sum_{a,b=xz, yz}\sum_{{\bf k};ss'}\ell^z_{ab}\ \sigma^z_{ss'} c^\dagger_{a{\bf k}s} c_{b{\bf k}s'}, +\ldots
\nonumber
\end{align}
where the first term reflects second-neighbor hopping 
the $xz$ and $yz$ orbitals, so 
$\lambda_{\bf k} \equiv 2\delta t \sin k_x \sin k_y$, and the second term represents the Ru atomic spin-orbit coupling 
where $ \ell^c_{ab} = i \epsilon_{a b c} $ are the spin-1 matrices 
representing the effective orbital angular momentum of the $t_{2g}$ orbitals and $ \sigma^a$ are the usual Pauli spin-matrices. 
We can see that if we include the $xy$ orbital, the atomic spin-orbit term will include the additional terms $\ell^x \sigma^x + \ell^y \sigma^y$. These terms, however, do not alter any of our qualitative results, except those we discuss in Section \ref{SOaction} and Appendix \ref{App:dLock}.
 
A more detailed model of the electronic structure of Sr$_2$RuO$_4$ might include additional terms in $\delta H$ (indicated by $\ldots$) 
but we will show that these terms are certainly smaller than the terms we have kept and therefore do not qualitatively effect the outcome of the calculations, so long as we focus on the limit
\begin{equation}
t \gg |\delta t| \sim |\eta| \gg |\Delta|.
\label{inequality}
\end{equation}
Eq.\eqref{EQ:orbHyb} omitted the interobital interaction terms of the form
\begin{equation}
\delta H_{int} = V \sum_{i \sigma \sigma' }d^{\dagger}_{xz,i \sigma} d^{\dagger}_{yz, i \sigma'} d_{yz, i \sigma'} d_{xz, i \sigma}
\end{equation}
that are local and therefore play a role everywhere in momentum space. However, in the BCS ground state, such interactions cause scattering only in the Cooper channel, and therefore play an appreciable role only in momentum points where the bands cross. Additional interaction terms such as the interorbital singlet-pair hopping terms do not play a role in the spin-triplet superconductor. Among higher order interaction processes, only those terms proportional to the interorbital susceptibility (vanishingly small in the present case) affect the interorbital modes. Thus, we are led to the same conclusion: the mixing between the different orbitals whether it be due to tunneling or to interactions, is weak.
 
\section{Derivation of the NLSM} 
 
In this section, we will compute the terms in the NLSM
in Eq.\ref{EQ:nlsm} from the microscopic model in Eqs. \ref{H}-\ref{EQ:orbHyb}. 

\subsection{Kinetic terms, ${\cal W}$ and ${\cal M}$}
Firstly, we derive expressions for the time-derivative terms in Eq.\ref{EQ:nlsm} which do not involve mixing of the orbitals or spin-orbit coupling; to leading order, these terms can be computed in the limit $\delta H=0$ and $\Delta_0=0$. Because the number density is canonically conjugate to the phase of the superconducting order, these terms ultimately reflect the energy cost of shifting electrons from one orbital to another or from one spin polarizing to the other. To quadratic order, the cost in energy per unit volume associated with a transfer of charge $\delta n$ per unit cell from one band to the other and with a spin density ${\bf S}_a$ in band $a$ is
\begin{equation}
K[\delta n, {\bf S}_a] = \frac{1}{4\chi_c}(\delta n)^2 + \frac{1}{2\chi_{sp}}\sum_a \left(\frac{{\bf S}_a}{\hbar}\right)^2,
\label{time}
\end{equation}
where 
$\chi_c = N(0)$ and $ \chi_{sp} = N(0)/4$ are the density and spin susceptibilities for each quasi-1D band.
In the standard fashion, spin-triplet Cooper pairing gives rises to the commutation relations \cite{LEGGETT1975, Leggett1974, Vollhardt1990}
\begin{align}
[\delta n, e^{i\phi_-}]&= -4ie^{i\phi_-},\nonumber\\
[S^i_a, \hat{d}^j_b] &= i\hbar\epsilon^{ijk} \delta_{ab}\hat{d}^k_b;
\label{EQ:conjugate}
\end{align}
thus, we can regard $K$ as the `kinetic' energy density of $\phi_-$ and ${\bf \hat{d}}_a$. By using the Heisenberg equations of motions $\partial_t \phi = i[\phi, K]/\hbar$ and $\partial_t {\bf \hat{d}}_a = i[{\bf \hat{d}}_a, K]/\hbar$, we obtain
\begin{equation}
K[\phi, {\bf \hat{d}}_a] = \frac{\hbar^2 \chi_c}{16}(\partial_t \phi)^2 + \frac{\hbar^2\chi_{sp}}{2}\sum_a (\partial_t {\bf \hat{d}}_a)^2.
\label{EQ:kinEff}
\end{equation}
This is equivalent to the first line of the NLSM action of Eq.\ref{EQ:nlsm}.

\subsection{Relative Josephson coupling, ${\cal J}$}
\begin{figure}
\includegraphics[width=1.25in]{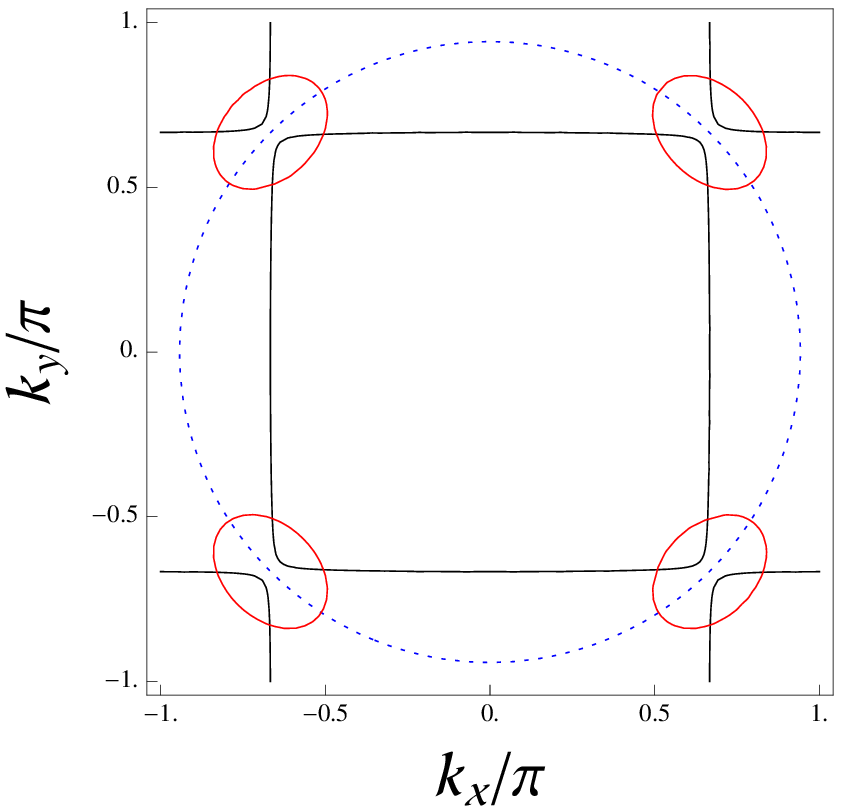}\includegraphics[width=2.1in]{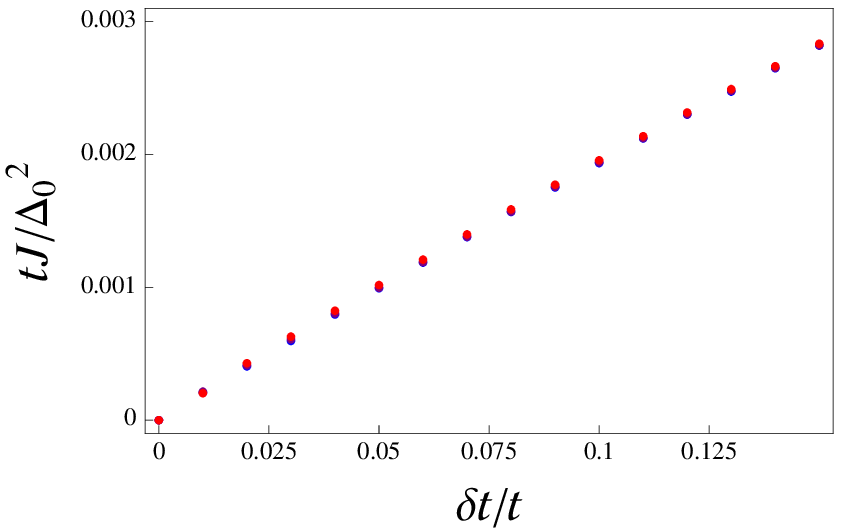}
\caption{The left panel shows as the black solid curves a schematic plot of the Fermi surface originating from the quasi-1D bands which derive primarily from the Ru $d_{xz}$ and $d_{yz}$ orbitals; it was computed from the the microscopic model of Section II. (The blue dotted curve represents the Fermi surface of the quasi 2D band arising from the Ru $d_{xy}$ orbitals, which are not included explicitly in the model.) The avoided crossings in the red circled regions reflect the (clearly small) effects of the orbital mixing terms on the Fermi surface structure.
The right panel shows the intercomponent Josephson coupling, ${\cal J}_0$, defined in Eq. \ref{calJ} as a function of 
the interorbital coupling energy, $\delta t$, computed by numerically performing the integral in Eqs. \ref{calJ0} and \ref{EQ:JBdG} in Appendix \ref{App:interBdG} 
with $\Delta_0/t = 0.002, 0.004, 0.008$ and $\eta=0$; the result confirms an approximate linear dependence of ${\cal J}_0$ on $|\delta t|\Delta_0^2$, as in Eq.\ref{EQ:interApprox}.}
\label{FIG:JSmall}
\end{figure}

The leading contribution to the interorbital Josephson coupling, $\mathcal{J}$, can be computed in the $\eta = 0$ limit. 
As a warm up, we first calculate the ground state energy for the spinless case where the relative phase between the condensate of two components of the order parameter is set to $\phi_-=\theta_x-\theta_y$. The ground state energy can be computed as
\begin{equation}
\mathcal{E}_0 (\phi_-) = -
\sum_{\alpha{\bf k}}[E_{\alpha{\bf k}}(\phi_-) 
-
{\xi}_{\alpha{\bf k}}]/2,
\end{equation}
where $E_{\alpha{\bf k}}(\phi_-)$'s are the 
eigenenergies of the spin ``up'' quasiparticles in $\mathcal{H}_{BdG}$ in Eq. \ref{EQ:1D0} with $\hat d_x = \hat d_y=\hat z$ and $\lambda=0$. %
As we show in Appendix \ref{App:interBdG}, the local stability of the chiral state 
follows from from the fact that 
$\mathcal{E}'_0 (\phi_- = \pm \pi/2)= 0$ and $\mathcal{E}''_0 (\phi_- = \pm \pi/2)>0$.
Indeed, since in the limit of weak mixing, we know the dependence of the energy on $\phi_-$ must be approximately harmonic, it follows that
$\mathcal{E}''_0$ is related to the value of the Josephson coupling. Time reversal symmetry implies that $\mathcal{E}_0 (\phi_-)= \mathcal{E}_0 (-\phi_-)$ and the fact that $\Delta_\alpha$ transforms like a vector under rotations by $\pi/2$ implies that $\mathcal{E}_0 (\phi_-)= \mathcal{E}_0 (\pi-\phi_-)$. Thus,
\begin{align}
\mathcal{E}_0 (\phi_-)
\approx \ {\rm const.} + {\cal J}_0 \cos (2\phi_-),
\label{EQ:minApprox}
\end{align}
where
\begin{equation}
{\cal J}_0 = \frac{1}{4}\mathcal{E}''_0 \left(\phi_-= \frac{\pi}{2}\right) = -\frac{1}{8}\sum_{\alpha{\bf k}}E''_{\alpha{\bf k}}\left(\phi_-= \frac{\pi}{2}\right).
\label{calJ0}
\end{equation}
We have computed ${\cal J}_0$ by numerically diagonalizing the model in Eq. \ref{H} with $\eta=0$ -- see Fig. \ref{FIG:JSmall} and Appendix \ref{App:interBdG}. 

It is straightforward to generalize the above result to the spinful case 
and obtain $\mathcal{J}$ of Eq.\eqref{EQ:nlsm}. From the definition of the $d$-vector \cite{Vollhardt1990, MACKENZIE2003}
\begin{equation}
\left[\begin{array}{cc}\Delta_{\alpha\uparrow\uparrow} & \Delta_{\alpha\uparrow\downarrow}\\ \Delta_{\alpha\downarrow\uparrow} & \Delta_{\alpha\downarrow\downarrow}\end{array}\right] \equiv \Delta f_\alpha({\bf k})e^{i\theta_\alpha}\hat d_\alpha \cdot\left[\begin{array}{cc}-\hat{e}_x + i\hat{e}_y & \hat{e}_z\\ \hat{e}_z & \hat{e}_x + i\hat{e}_y\end{array}\right],
\label{EQ:dDef}
\end{equation}
it follows that it is always possible to chose the spin quantization axis, $\hat e_z$, to be perpendicular to both $\hat{d}_{x}$ and $\hat{d}_{y}$,
in which case all pairing is between like spins ($\Delta_{\alpha\uparrow\downarrow}=0$).
Thus, for $\eta=0$, the Josephson coupling is a sum of terms two equal contributions from spin-up and spin-down pairs: 
\begin{equation}
\mathcal{E}_0 (\phi_{\uparrow\!\uparrow},\!\phi_{\downarrow\!\downarrow})
\!=\!{\cal J}_0(\cos 2\phi_{\uparrow\!\uparrow}\!+\!\cos 2\phi_{\downarrow\!\downarrow})
\label{EQ:spinJ}
\end{equation}
where $\theta_{\uparrow\uparrow} = \theta_x - \theta_y +\alpha+\pi$, $\theta_{\downarrow\downarrow} = \theta_x - \theta_y -\alpha$ ,
and $\cos \alpha=\hat d_x\cdot\hat d_y$. 
A small exercise in trigonometry thus leads to from this result to Eq. \ref{calJ} with Eq. \ref{calJ0} for ${\cal J}_0$.
 
We have numerically evaluated the integral over ${\bf k}$ in Eq. \ref{calJ0} to obtain the results for ${\cal J}_0$ shown in Fig.\ref{FIG:JSmall}.
It is clear that ${\cal J}_0$ scales as
\begin{equation}
{\cal J}_0\sim \frac{|\delta t|\ \Delta_0^2}{t^2} \sim \frac{|\delta t|}{t} N(0) \Delta_0^2,
\label{EQ:interApprox}
\end{equation}
where $N(0)=16\pi/(\sqrt{3}t)$ is the density of states for each quasi-1D orbital; this can be also obtain through analytic approximation presented in Appendix \ref{App:interBdG}. As promised, ${\cal J}_0$ is parametrically smaller than the condensation energy per unit volume, $N(0)\Delta_0^2/2$.
 
A few points are worth noting. Firstly, this result is non-analytic in $\delta t$, from which one concludes that it is non-perturbative. In Appendix \ref{App:perturb}, we compute ${\cal J}_0$ perturbatively in powers of $\delta t$, in which limit we obtain ${\cal J}_0 \sim N(0) (\delta t)^4/(t\Delta_0)$, which is an analytic function of $\delta t$, but non-analytic in $\Delta_0$. This reflects the fact that the perturbative expression is valid only for $|\delta t| \ll \Delta_0$, a physically unreasonable restriction. The perturbative expression does, however, match smoothly to the non-perturbative one when $|\delta t| \sim \Delta_0$. The origin of Eq. \ref{EQ:interApprox} can be understood intuitively as arising from the quasi 1D character of the bands. The contribution to the condensation energy from the portions of the bands away from the crossing points (enclosed by red circles in Fig. \ref{EQ:interApprox}) is largely insensitive to orbital mixing - only in a neighborhood of width $\sim |\delta t/t|$ about the crossing points is orbital mixing significant, but there it makes an $\mathcal O(1)$ change in the condensation energy. These considerations lead to the proposed scaling expression.
On the other hand, if the chiral pairing originated from the nearly circular 2D Fermi surface, 
changes in the relative phase 
of the $p_x$ and the $p_y$ 
components of the pair-field affect the pairing gap magnitude 
over most of the Fermi surface, so ${\cal J}_0$ (or more properly, $\mathcal{E}''_0$) must be order of the total condensation energy \cite{WOLFLE1976, Volovik1992, Higashitani2000}.
 
\subsection{Spin-orbit term, $\Gamma_0$}
\label{SOaction}
  
The $xy$ orbital does not play a direct role in our analysis, but does contribute to the intra-orbital $d$-vector locking $\Gamma$. While we will not consider the superconductivity in this orbital, 
it does affect the pairing interaction in the quasi-1D orbitals. It is due to the spin-orbit coupling involving the $xy$ orbitals that we have anisotropy in the spin channel of the pairing interaction. More specifically, in both quasi-1D orbitals, the effective pairing interaction for $\hat{d} \parallel \hat{z}$ will be stronger than the effective pairing interaction for $\hat{d} \perp \hat{z}$. This difference in the pairing interaction can be estimated as
$\delta V/V \propto (\eta/t)^2$
as it is due to the electrons having intermediate states in the $xy$ orbitals (see Appdix \ref{App:dLock} for derivation). Now, from the BCS self-consistency condition $\log (t/\Delta) \propto t/V$, the pairing gap will change by
\begin{equation}
\frac{\delta \Delta}{\Delta} \propto \frac{t\delta V}{V^2} = \left(\frac{\eta}{t}\right)^2 \log \frac{t}{\Delta}
\end{equation}
for small $\delta V$. 
Therefore, the condensation energy for the case $\hat{d} \parallel \hat{z}$ should be slightly larger by
\begin{equation}
\Gamma_0 \equiv \frac{1}{2}N(0)\delta(\Delta^2) \propto \left[\left(\frac{\eta}{t}\right)^2 \log \frac{t}{\Delta}\right] N(0)\Delta^2
\label{EQ:dVecPin}
\end{equation}
than that of the case $\hat{d} \perp \hat{z}$; 
this is the origin of the intra-orbital $d$-vector locking 
in Eq.\ref{EQ:nlsm}. 
We emphasize that the energy scale of this locking is much smaller than that of the condensation energy, which is consistent with the $c$-axis Knight shift experiment \cite{MURAKAWA2004}. \footnote{In terms of the Zeeman field $H_{rot}$ which rotates the $d$-vector into the $ab$-plane, we have $\Gamma_0 = \chi_{sp}(\mu_B H_{rot})^2/2$\cite{Vollhardt1990}.}
However, this small $d$-vector locking, unlike the small $J$, is not a characteristic unique to the quasi-1D model, as the pairing interaction anisotropy due to the same physics can also occur in the 2D orbital.
 
 
\section{Normal modes}
In the usual fashion, we can obtain an understanding of the low energy collective modes from the equations of motion 
derived from the non-linear sigma model \cite{Leggett1974} in Eq. \ref{EQ:nlsm}.
The $d$-vector dynamics 
are those of coupled pendulums, while the relative phase, $\phi_-$ executes familiar Josephson oscillations. 
Looking at these modes in the limit that ${\bf k}\to {\bf 0}$
(spatially homogeneous modes) and for small amplitude deviations from the A-phase ground-state, in which $\phi_-=0$ and $\hat d_x=\hat d_y=\hat z$, we find
\begin{align}
\frac{\hbar^2 \chi_c}{16} \partial^2_t \phi_- =& 4{\cal J}_0\phi_-,\nonumber\\
\frac{\hbar^2\chi_{sp}}{2}\partial^2_t\left[\begin{array}{c} \delta{\bf \hat{d}}_x\\ \delta{\bf \hat{d}}_y\end{array}\right] =& -\left[\begin{array}{cc}\Gamma_0 + 4{\cal J}_0 & -4{\cal J}_0\\ -4{\cal J}_0 & \Gamma_0+4{\cal J}_0\end{array}\right]\left[\begin{array}{c} \delta{\bf \hat{d}}_x\\ \delta{\bf \hat{d}}_y\end{array}\right].
\end{align}
where $\delta{\bf \hat d}_a\cdot \hat e_z=0$.
The above equation of motion tells is that the small value of $J$ in the quasi-1D model gives us more soft collective modes than the 2D model. 
From this we can deduce the the gaps (or ``masses'') of three distinct normal modes, all of which vanish in the limit $\delta H\to 0$, {\it i.e.} at the multcritical point: 
\begin{align}
m_c &= 8 \sqrt{{\cal J}_0/\chi_c} = ( \gamma_c)\Delta_0,\nonumber\\
\label{mass}
m_{s+} &= \sqrt{{2\Gamma_0}/{\chi_{sp}}}= (\gamma_{s+}) \Delta_0,\\
m_{s-} &= \sqrt{{2(\Gamma_0+8{\cal J}_0)}/{\chi_{sp}}}= (\gamma_{s-}) \Delta_0,\nonumber,
\end{align}
where for the relative phase mode, $\gamma_c\sim |\delta t|/t$, for the in-phase spin-wave mode, $\gamma_{s+}\sim (\eta/t)^2$, while for the relative d-vector orientation mode, $\gamma_{s-}\sim$ the larger of $\gamma_c$ and $\gamma_{s-}$.
We also see, as is natural, that the in-phase spin-wave mode is unaffected by the interorbital coupling.

\section{Discussion}

Most studies to date have worked with the assumption that $\gamma$ is the active band.  From an experimental perspective, heat capacity measurements\cite{Deguchi2004} showed that the fraction of density of states at the Fermi level that was depleted at T$_c$ is consistent with the contribution from the $\gamma$ band.  However, the balance is delicate, as the contribution from $\{ \alpha, \beta\}$ to the total density of states is similar to the contribution from  $\gamma$.  From a theoretical perspective, asymptotically exact  calculations in the weak-coupling limit involving all three Fermi surfaces lead  inevitably to the conclusion that the dominant pairing strength occurs among the 1D bands\cite{RAGHU2010}. However, when stronger electron correlations are present, the validity of these results is unclear. In this regard, it is important to consider experimental signatures that may help to discriminate between the two possibilities for the active orbitals.  This has been the primary motivation for carrying out the present analysis.  
 
We have shown 
that if the chiral p-wave superconductivity in Sr$_2$RuO$_4$ arises from the quasi-1D bands, this implies that the collective properties are controlled by the existence of a nearby multicritical point at which there is an enlarge emergent order parameter symmetry and correspondingly a set of anomalously soft ``almost Goldstone'' soft collective modes.
 
Because of the relatively weak spin-orbit coupling in Sr$_2$RuO$_4$, the in-phase spin-wave mode is expected to have energy small compared to the superconducting gap. An estimate of the in-phase spin-wave gap can be obtained on the basis of experiments on the the $c$-axis Knight shfit\cite{MURAKAWA2004}, from which it follows that $\gamma_{s+}=m_{s+}/\Delta_0$ is less than 0.01.
This result implies an extremely small value of $\Gamma_0$ in Eq. \ref{EQ:nlsm}, but does not distinguish between different microscopic origins of the pairing. We have therefore focused,
in particular, on the gap (mass) of the relative phase and relative d-vector orientational modes, which are analogues of the ``clapping modes'' familiar from studies of $^3$He-A \cite{WOLFLE1976, Volovik1992} and also investigated in the chiral $d$-wave superconductor \cite{BALATSKY2000}. The corresponding mode frequencies have been computed in the context of the quasi-2D bands, leading to the prediction that they would have an energy $\sqrt{2}\Delta_0$ ($\gamma_c=\gamma_{s-}=\sqrt{2}$)\cite{Higashitani2000, Kee2000}.
We recover their result from our NLSM if we extrapolate the results to the case of strong inter-component Josephson coupling, where ${\cal J}_0$ is comparable to the condensation energy density. However, because inter-orbital mixing is relatively weak, the corresponding modes are expected to have parametrically lower energy if the superconductivity arises in the quasi 1D bands - hence Eq. \ref{mass}. (For more on this correspondence see Appendix \ref{App:2Dmodel}.)
 
The various collective modes in Sr$_2$RuO$_4$ can, in principle be detected using methods that have already been discussed in the literature for the 2D model, which includes, among others, electron spin resonance \cite{Tewordt1999, MIYAKE2010} (see Appendix \ref{App:detect}), ultrasound attenuation and Raman scattering \cite{Kee2000}. However, we can 
expect the inter-orbital `nearly Goldstone' modes to have much lower energy if the pairing originates primarily on the quasi 1D bands, than if it arises on the quasi 2D band.
 
 
Our analysis has potential implications for various existing experimental puzzles 
concerning the properties of Sr$_2$RuO$_4$ near T$_c$. If the phase transition exhibits mean-field behavior, an inescapable consequence is that the transition must be split by a field applied in the basal plane. The fact that this does not occur in Sr$_2$RuO$_4$
suggests that fluctuations may play a significant role in the transition. The higher emergent symmetry of the proximate multicritical point would be an obvious source of anomalously strong fluctuations. In this regard it is worth noting that in two dimensions, a non-zero (Kosterliz-Thouless) transition is still possible, despite the existence of gapless spin-wave like modes, but for an order parameter with a larger continuous symmetry, such fluctuations necessarily reduce the transition temperature to $T=0$, or to a low temperature at which explicit symmetry breaking terms, or three dimensional couplings cut-off these fluctuations.
These issues will be investigated in depth in a future publication.
 
{\bf Aknowledgements:} We thank E.~Abrahams, S.~Brown, S.~Chakravarty, A.~Chubukov, E.~Fradkin, L.~Fu, T.~Geballe, C.~Kallin, H.-Y.~Kee, D.~J.~Scalapino and B.~Spivak 
for helpful discussions. This work was supported in part by DE-AC02-76SF00515 (SBC, AK, and SAK), 
and the Alfred P. Sloan Foundation (SR). We thank the Aspen Center for Physics and the KITP, where part of this work was carried out, for hospitality during the workshop `New Topological States of Quantum Matter' and the program `Topological Insulators and Superconductors' respectively.
 
\bibliography{collective_final}
 
\appendix
  
\begin{widetext}
\section{Derivation of inter-orbital Josephson coupling}
 
\subsection{BdG formalism}
\label{App:interBdG}
 
Since the sum of the quasi-particle eigenenergy gives us the total ground state energy, in obtaining the ground state energy a given relative phase $\phi$ between two orbitals, the most straightforward method is through calculating the BdG eigenenergies when the relative phase is $\phi$. 
We can compute these eigenenergies from diagonalizing the spinless BdG Hamiltonian,
\begin{equation}
H_{BdG} (\phi_-) =\left[\begin{array}{cccc} \xi_{x{\bf k}} & \lambda_{\bf k} & \Delta_0 \sin k_x & 0\\
\lambda_{\bf k} &\xi_{y{\bf k}} & 0 & e^{i\phi_-} \Delta_0 \sin k_y\\
\Delta_0 \sin k_x & 0 & -\xi_{x{\bf k}} & -\lambda_{\bf k}\\
0 & e^{-i\phi_-} \Delta_0 \sin k_y & -\lambda_{\bf k} & -\xi_{y{\bf k}}\end{array}\right],
\end{equation}
where our basis is $(u_{x{\bf k}}, u_{y{\bf k}}, v_{x{\bf k}}, v_{y{\bf k}})^T$. This gives us the eigenenergies of
\begin{equation}
E_{{\bf k}\pm}(\phi_-) = \frac{1}{\sqrt{2}}\sqrt{A_{\bf k} \pm \sqrt{B_{\bf k}(\phi_-)}},
\end{equation}
where
\begin{align}
A_{\bf k} \equiv& \xi^2_{x{\bf k}} + \xi^2_{y{\bf k}} + \Delta_0^2 (\sin^2 k_x + \sin^2 k_y)+2\lambda^2_{\bf k},\nonumber\\
B_{\bf k}(\phi_-) \equiv& [(\xi^2_{x{\bf k}} -\xi^2_{y{\bf k}})+\Delta_0^2 (\sin^2 k_x -\sin^2 k_y)^2]^2 + 4\lambda_{\bf k}^2[(\xi_{x{\bf k}}+\xi_{y{\bf k}})^2 + \Delta_0^2 (\sin^2 k_x + \sin^2 k_y)]\nonumber\\
&- 8\lambda_{\bf k}^2 \Delta_0^2 \sin k_x \sin k_y \cos \phi_-.
\label{EQ:eigenPi}
\end{align}
 
The ground state energy has minima at $\phi_- = \pm \pi/2$ and therefore can be approximated by Eq.\ref{EQ:minApprox}.
To see this note that the total ground state energy 
\begin{equation}
\mathcal{E}_0 (\phi_-) = {\rm const} -\frac{1}{2}\sum_{\bf k} 
[E_{{\bf k}+}(\phi_-)+E_{{\bf k}-}(\phi_-)].
\end{equation}
when differentiated by $\phi$ gives us
\begin{equation}
\mathcal{E}_0'(\phi_-)= \sin\phi_-\sum_{\bf k} \frac{\lambda_{\bf k}^2 \Delta_0^2 \sin k_x \sin k_y}{E_{{\bf k}+}(\phi_-)E_{{\bf k}-}(\phi_-)[E_{{\bf k}+}(\phi_0)+E_{{\bf k}-}(\phi_-)]}.
\end{equation}
for which we used
\begin{equation}
E'_{{\bf k}\pm}(\phi_-) = \pm\frac{2\lambda^2_{\bf k} \Delta_0^2 \sin k_x \sin k_y}{E_{{\bf k}\pm}(\phi_-)[E^2_{{\bf k}+}(\phi_-)-E^2_{{\bf k}-}(\phi_-)]}\sin \phi_-.
\end{equation}
This derivative vanishes at $\phi = \pm \pi/2$, as $E_{{\bf k}\pm}(\phi = \pm \pi/2)$'s are even in $k_x$ and $k_y$. 
From the second derivative
\begin{equation}
\mathcal{E}_0''(\phi_-=\pi/2)
= 
2\sum_{\bf k}\frac{\lambda^4_{\bf k}\Delta_0^4 \sin^2 k_x \sin^2 k_y}{(E_{{\bf k}+}E_{{\bf k}-})^2(E_{{\bf k}+}+E_{{\bf k}-})}\left[\frac{1}{E_{{\bf k}+}E_{{\bf k}-}}+\frac{1}{(E_{{\bf k}+}+E_{{\bf k}-})^2}\right]_{\phi_- = \pm \frac{\pi}{2}},
\label{EQ:JBdG}
\end{equation}
which is clearly positive we see that these extrema at $\phi_- = \pm \pi/2$ are minima. Lastly, we note that $\mathcal{E}_0 (\phi_-)$ is $\pi$-periodic in $\phi_-$, as Eq.\ref{EQ:eigenPi} shows that each eigenenergy is invariant under the combination of $\phi_- \to \pi-\phi_-$ and $\pi/2$ rotation in the $k$-space.

We now need to evaluate Eq.\eqref{EQ:JBdG}. 
Since the main contributions will come from the four crossing points of the 1D orbitals, we take the following expansion around $(k_x,k_y) = (2\pi/3, 2\pi/3)$:
\begin{equation}
E_\pm = \sqrt{\left[\frac{v(k_x+k_y-4\pi/3)}{2}\pm\sqrt{\frac{v^2(k_x-k_y)^2}{4} + \lambda^2}\right]^2+\bar{\Delta}^2} = \lambda \sqrt{\left(q_x \pm \sqrt{q_y^2+1}\right)^2+\frac{\bar{\Delta}^2}{\lambda^2}},
\end{equation}
where $v = \sqrt{3}t/2$, $\lambda = 3|\delta t|/2$, $\bar{\Delta} = \Delta_0 |\sin 2\pi/3|$, and $q_x = v(k_x+k_y-4\pi/3)/2\lambda, q_y = v(k_x-k_y)/2\lambda$. We also note the following two points: i) ${\rm min}(E_\pm) = \bar{\Delta}$ and ii) when $E_+$ ($E_-$) is at its minimum, $E_{-(+)} \sim \lambda \gg \bar{\Delta}$ so ${\rm min}(E_+ + E_-) \sim \lambda$. This leads to the following approximation:
\begin{align}
{\cal J}_0 = \frac{1}{4}\mathcal{E}_0''(\phi_-=\pi/2) \approx & \frac{1}{2}\sum_{\bf k}\frac{\lambda^4_{\bf k}\Delta_0^4 \sin^2 k_x \sin^2 k_y}{(E_{{\bf k}+}E_{{\bf k}-})^3(E_{{\bf k}+}+E_{{\bf k}-})} = \frac{\lambda^4\bar{\Delta}^4}{8\pi^2}\frac{2\lambda^2}{v^2}\int d^2 q \frac{1}{(E_{{\bf k}+}E_{{\bf k}-})^3(E_{{\bf k}+}+E_{{\bf k}-})}\nonumber\\
\approx& \frac{2\bar{\Delta}^4}{\pi^2\lambda v^2} \int d^2 q \left[\left(q_x + \sqrt{q_y^2+1}\right)^2+\frac{\bar{\Delta}^2}{\lambda^2}\right]^{-\frac{3}{2}}\left[\left(q_x - \sqrt{q_y^2+1}\right)^2+\frac{\bar{\Delta}^2}{\lambda^2}\right]^{-3}\nonumber\\
\approx&\frac{\bar{\Delta}^4}{8\pi^2\lambda v^2} \int d \tilde{q}_x dq_y \frac{1}{(\tilde{q}_x^2 + \bar{\Delta}^2/\lambda^2)^{\frac{3}{2}}(q_y^2+1)^2} = \frac{\bar{\Delta}^4}{8\pi^2\lambda v^2}\frac{2\lambda^2}{\bar{\Delta}^2}\frac{\pi}{2}\nonumber\\
=& \frac{3}{16\pi}\frac{|\delta t|}{t}\frac{\Delta_0^2}{t}.
\end{align}
We obtained the same dependence on parameters as in Fig.\ref{FIG:JSmall}, though the coefficient came out about an order of magnitude larger.
 
\subsection{Perturbation method}
\label{App:perturb}
 
We show here that once we ignore the spin-orbit coupling, Due to the $C_4$ symmetry, this inter-orbital Josephson coupling is zero for the lowest order. To see this, we note that from the second order perturbation theory
\begin{equation}
\mathcal{E}^{(1)}_J (\phi_-) = \sum_{{\bf k}s}\frac{\langle \lambda_{\bf k}c^\dagger_{x{\bf k}s}c_{y{\bf k}s} \lambda_{-{\bf k}} c^\dagger_{x,-{\bf k}s}c_{y,-{\bf k}s} + {\rm h.c.}\rangle}{-E_{x{\bf k}} - E_{y{\bf k}}}
= -(\delta t)^2 \sum_{{\bf k}s}\frac{\sin^2 k_x \sin^2 k_y}{E_{x{\bf k}} + E_{y{\bf k}}}\frac{\Delta^*_{x{\bf k};ss}}{E_{x{\bf k}}}\frac{\Delta_{y{\bf k};ss}}{E_{y{\bf k}}} + {\rm c.c.} = 0,
\end{equation}
where $E_{a{\bf k}} \equiv \sqrt{\xi^2_{a{\bf k}} + |\Delta_{a{\bf k}}|^2}$, and we used
\begin{equation}
\langle c_{a,-{\bf k},s'} c_{a,{\bf k},s}\rangle = -\frac{\Delta_{a{\bf k};ss'}}{2E_{a{\bf k}}}; 
\end{equation}
this result is basically due to $\Delta_{x(y){\bf k}}$ being odd in $\sin k_{x(y)}$.
 
Therefore, it is from the second order inter-orbital Josephson coupling that gives rise to the dependence of the energy on the relative phase $\phi$ and the spin state $\hat{d}_{xz,yz}$. Given that we have completely decoupled opposite spins, we only need to consider the process that tunnels two spin up-up pairs and two spin down-down pairs. This can be calculated from the fourth order perturbation theory:
\begin{align}
\mathcal{E}^{(2)}_J (\phi_-) =& \sum_{{\bf k}'{\bf k};s}\langle \lambda_{\bf k}c^\dagger_{x{\bf k}s}c_{y{\bf k}s} \lambda_{{\bf k}'} c^\dagger_{x{\bf k}'s} c_{y{\bf k}'s} \lambda_{-{\bf k}} c^\dagger_{x,-{\bf k}s}c_{y,-{\bf k}s} \lambda_{-{\bf k}',s} c^\dagger_{x,-{\bf k}',s} c_{y,-{\bf k}',s} + {\rm h.c.}\rangle 
\nonumber\\
&\times \frac{1}{(-E_{x{\bf k}} - E_{y{\bf k}})}\frac{1}{(-E_{x{\bf k}} - E_{y{\bf k}}-E_{x{\bf k}'} - E_{y{\bf k}'})}\left(\frac{1}{-E_{x{\bf k}} - E_{y{\bf k}}}+\frac{1}{-E_{x{\bf k}'} - E_{y{\bf k}'}}\right)\nonumber\\
=& \sum_{{\bf k}'{\bf k};s}\lambda_{\bf k}\lambda_{-{\bf k}}\lambda_{{\bf k}'}\lambda_{-{\bf k}'}\left[\left(\frac{\Delta^*_{x{\bf k};ss}}{2E_{x{\bf k}}}\right)\left(\frac{\Delta_{y{\bf k};ss}}{2E_{y{\bf k}}}\right)\left(\frac{\Delta^*_{x{\bf k}';ss}}{2E_{x{\bf k}'}}\right)\left(\frac{\Delta_{y{\bf k}';ss}}{2E_{y{\bf k}'}}\right)+{\rm c.c.}\right](1-\delta_{{\bf k}'{\bf k}})(1-\delta_{{\bf k}',-{\bf k}})\nonumber\\
&\times \frac{1}{(-E_{x{\bf k}} - E_{y{\bf k}})}\frac{1}{(-E_{x{\bf k}} - E_{y{\bf k}}-E_{x{\bf k}'} - E_{y{\bf k}'})}\left(\frac{1}{-E_{x{\bf k}} - E_{y{\bf k}}}+\frac{1}{-E_{x{\bf k}'} - E_{y{\bf k}'}}\right)\nonumber\\
=& \sum_{{\bf k}s}\frac{\lambda_{\bf k}^4/8}{(E_{x{\bf k}} + E_{y{\bf k}})^3}\frac{[(\Delta^*_{x{\bf k}';ss}\Delta_{y{\bf k}';ss})^2 + {\rm c.c.}]}{(E_{x{\bf k}} E_{y{\bf k}})^2},
\end{align}
which gives us
\begin{equation}
{\cal J}_0 = \sum_{\bf k}\frac{\lambda_{\bf k}^4/4}{(E_{x{\bf k}} + E_{y{\bf k}})^3}\frac{\Delta_0^4 (\sin k_x \sin k_y)^2}{(E_{x{\bf k}} E_{y{\bf k}})^2} \sim \frac{(\delta t)^4}{t^2 \Delta}.
\end{equation}
Note that this result is consistent with our BdG calculation, for it is qualitatively the same as taking the first term of Eq.\eqref{EQ:JBdG} which is much larger than the second term in the $|\Delta| \gg |\delta t|$ limit.
 
\section{$d$-vector locking}
\label{App:dLock}
 
To see why we need the spin involving the $xy$ orbital to lock the $d$-vector along the $c$-axis, we need to examine the spin-orbit coupling part - with the $xy$ orbital included - of the orbital hybridization of Eq.\eqref{EQ:orbHyb}:
\begin{equation}
\delta H_{kin} = \eta\sum_{a,b}\sum_{{\bf k};ss'}\bm \ell_{ab}\cdot {\bm \sigma}_{ss'} c^\dagger_{a{\bf k}s} c_{b{\bf k}s'}.
\end{equation}
This means that when we only include the spin-orbit coupling between the 1D orbitals, $S_z$ will remain a good quantum number while $S_{x,y}$ will not. Therefore, it is energetically more favorable to have an equal-spin pairing with the spin quantization axis along the $z$ direction, which equivalent to the $d$-vector lying in the $xy$ plane. For the same reason, the spin orbit coupling between the $xz$ and $xy$ orbital will favor the $d$-vector lying in the $yz$ plane and that between the $yz$ and $xy$ orbital the $d$-vector lying in the $xz$ plane. We therefore conclude that the $d$-vector is locked to the $c$-axis because the spin-orbit couplings involving the $xy$ orbital have larger effect than the spin-orbit coupling involving only the 1D orbitals.
 
We will now show that the pairing interaction anisotropy in the normal state is proportional to $\eta^2$. To do so, we will calculate how the normal state pair-field susceptibility, which is proportional to the normal state pairing interaction. To account for the effect of the spin-orbit coupling, what we will calculate is the inter-orbital susceptibility of the intra-orbital pairs. Since the form of the spin-orbit coupling is the same for any pair of orbitals, we can expect the dependence on $\eta$ to be the same. Therefore, we will only look at the triplet pair susceptibility involving one pair in the $xz$ orbital and another pair on the $yz$ orbital:
\begin{equation}
\chi_{tSC;x-y}^{\bf \hat{d}}({\bf k},i\Omega) = \sum_{\alpha\beta\gamma\lambda}\int_0^\beta d\tau e^{i\Omega\tau}\langle T_\tau c^\dagger_{x,{\bf k},\alpha}(\tau) c^\dagger_{x,-{\bf k},\beta}(\tau)y_{x,-{\bf k},\gamma}(0) c_{y,{\bf k},\lambda}(0)\rangle (i\sigma_2 {\bf \hat{d}}\cdot {\bm \sigma})_{\alpha\beta}(-i {\bf \hat{d}}\cdot {\bm \sigma}\sigma_2)_{\gamma\lambda},
\end{equation}
ignoring the $xy$ orbital. In this case $S_z$ is a good quantum number, so we obtain
\begin{align}
\chi_{tSC;x-y}^\parallel({\bf k},i\Omega) =& \frac{1}{\beta}\sum_{i\omega_n}\sum_{\sigma=\uparrow,\downarrow}G_{y\sigma;x\sigma}(-{\bf k},-i\omega_n - i\Omega)G_{y\sigma;x\sigma}({\bf k},i\omega),\nonumber\\
\chi_{tSC;x-y}^{\bf \hat{z}}l({\bf k},i\Omega) =& \frac{1}{\beta}\sum_{i\omega_n}\sum_{\sigma=\uparrow,\downarrow}G_{y\sigma;x\sigma}(-{\bf k},-i\omega_n - i\Omega)G_{y\bar{\sigma};x\bar{\sigma}}({\bf k},i\omega),
\end{align}
for the $d$-vector in and out of plane, respectively. When we take into account that the Green function is diagonal not in the orbital basis but in the band basis, we can write
\begin{align}
\chi_{tSC;x-y}^\parallel({\bf k},i\Omega) =& \frac{1}{\beta}\sum_{i\omega_n}\sum_{\sigma=\uparrow,\downarrow}\sum_{\mu\nu}\frac{\langle y,-{\bf k},\sigma|\mu,-{\bf k},\sigma\rangle\langle \mu,-{\bf k},\sigma|x,-{\bf k},\sigma\rangle}{-i\omega_n-i\Omega-\tilde{\xi}_\mu (-{\bf k})}\frac{\langle y,{\bf k},\sigma|\nu,{\bf k},\sigma\rangle\langle \nu,{\bf k},\sigma|x,{\bf k},\sigma\rangle}{i\omega_n-\tilde{\xi}_\nu ({\bf k})},\nonumber\\
\chi_{tSC;x-y}^{\bf \hat{z}}({\bf k},i\Omega) =& \frac{1}{\beta}\sum_{i\omega_n}\sum_{\sigma=\uparrow,\downarrow}\sum_{\mu\nu}\frac{\langle y,-{\bf k},\sigma|\mu,-{\bf k},\sigma\rangle\langle \mu,-{\bf k},\sigma|x,-{\bf k},\sigma\rangle}{-i\omega_n-i\Omega-\tilde{\xi}_\mu (-{\bf k})}\frac{\langle y,{\bf k},\bar{\sigma}|\nu,{\bf k},\bar{\sigma}\rangle\langle \nu,{\bf k},\bar{\sigma}|x,{\bf k},\bar{\sigma}\rangle}{i\omega_n-\tilde{\xi}_\nu ({\bf k})},
\end{align}
where $\mu,\nu$ are band indices and $\tilde{\xi}_{\bf k}$'s are the normal state eigenenergies. Using the fact that $\sum_\mu \langle y{\bf k}|\mu{\bf k}\rangle \langle \mu{\bf k}|x,{\bf k}\rangle=0$ and, in an appropriate basis, $\langle a\bar{\sigma}|\mu\bar{\sigma}\rangle = \langle a\sigma|\mu\sigma\rangle^*$, $a$ being the orbital label, we obtain the anisotropy
\begin{align}
\Delta \chi_{tSC;x-y}({\bf k},i\Omega) \equiv& \chi_{tSC;x-y}^{\bf \hat{z}}({\bf k},i\Omega) - \chi_{tSC;x-y}^\parallel({\bf k},i\Omega)\nonumber\\ =&
\frac{2}{\beta}\left[|\langle y{\bf k}\uparrow|\alpha{\bf k}\uparrow\rangle \langle\alpha{\bf k}\uparrow|x{\bf k}\uparrow\rangle|^2-{\rm Re}(\langle y{\bf k}\uparrow|\alpha{\bf k}\uparrow\rangle \langle\alpha{\bf k}\uparrow|x{\bf k}\uparrow\rangle)^2\right]\nonumber\\
\times&\sum_{i\omega_n}\left[\frac{1}{-i\omega_n-i\Omega-\tilde{\xi}_\alpha (-{\bf k})}-\frac{1}{-i\omega_n-i\Omega-\tilde{\xi}_\beta (-{\bf k})}\right]\left[\frac{1}{i\omega_n-\tilde{\xi}_\alpha ({\bf k})}-\frac{1}{i\omega_n-\tilde{\xi}_\beta ({\bf k})}\right]
\end{align}
It is the transformation matrix between the orbital and the band basis that gives rise to the $\eta^2$ dependence.
 
To see this, note that the first quantized form of the normal state Hamiltonian can be written as
\begin{equation}
h_{kin} + \delta h_{kin} = \frac{\xi_x+\xi_y}{2} + \frac{\tilde{\xi}_\alpha -\tilde{\xi}_\beta}{2}
\left[\begin{array}{cc}\cos\rho & i\sigma\sin\rho\\ -i\sigma\sin\rho & -\cos\rho\end{array}\right]
\end{equation}
when ignoring the spin conserving orbital hybridization, with $\tan\rho = 2\eta/(\xi_x -\xi_y)$. This gives us $\langle y\uparrow|\alpha\uparrow\rangle = e^{-i\pi/4}\sin(\rho/2)$ and $\langle \alpha\uparrow|x\uparrow\rangle = e^{-i\pi/4}\cos(\rho/2)$, so we obtain
\begin{equation}
|\langle y\uparrow|\alpha\uparrow\rangle \langle\alpha\uparrow|x\uparrow\rangle|^2-{\rm Re}(\langle y\uparrow|\alpha\uparrow\rangle \langle\alpha\uparrow|x\uparrow\rangle)^2 = 2\cos^2(\rho/2)\sin^2(\rho/2) = \frac{\eta^2/2}{(\xi_x-\xi_y)^2/4+\eta^2},
\end{equation}
and the pair-field anisotropy at the zero temperature and frequency is
\begin{equation}
\left.\Delta \chi_{tSC;x-y}({\bf k},i\Omega=0)\right\vert_{T=0} = -\frac{1}{2}\frac{\eta^2}{[\eta^2-\xi_x ({\bf k}) \xi_y ({\bf k})][\xi_x ({\bf k}) +\xi_y ({\bf k})]}.
\end{equation}
\end{widetext}
 
\section{Experimental detection}
\label{App:detect}
 
In detecting the collective modes, we can either try to see the resonant response to some effective driving force or other excitations of Sr$_2$RuO$_4$ decay into the collective mode.
 
We will first discuss the resonant response. Any AC field that couples linearly to our collective modes can serve as a driving force. One well known case is the AC Zeeman field \cite{LEGGETT1975}, which couples through the Zeeman energy term $\mathcal{H}_Z = -(\mu_B/\hbar) {\bf H}\cdot \sum_a {\bf S}_a$ (where $\mu_B$ is the Bohr magneton). To see this, note that we can approximate, for a small $H$, ${\bf S}_a \approx \chi_{sp} {\bf \hat{z}} \times (\partial_t \hat{\bf d}_a)$, giving us the spin equations of motion
\begin{align}
&\frac{\chi_{sp}}{2}\partial^2_t \left[\begin{array}{c} \delta{\bf \hat{d}}_x\\ \delta{\bf \hat{d}}_y\end{array}\right] + \left[\begin{array}{cc}\Gamma + 4J & -4J\\ -4J & \Gamma+4J\end{array}\right]\left[\begin{array}{c} \delta{\bf \hat{d}}_x\\ \delta{\bf \hat{d}}_y\end{array}\right]\nonumber\\
=&-\frac{2\mu_B\chi_{sp}}{\hbar} \hat{z} \times \partial_t{\bf H}\left[\begin{array}{c} 1 \\ 1\end{array}\right],
\end{align}
which are just those of a pair of driven coupled harmonic oscillators. The sharp difference between the quasi-1D and the 2D model is that the former has a double resonance peak whereas the latter has only a single peak just like that of the longitudinal NMR in $^3$He-A \cite{OSHEROFF1972}.
The charge analogue for this would be a uniaxial AC strain along the $a(b)$-axis. This will give rise to a chemical potential difference between the $xz, yz$ orbitals, we expect the energy cost to be proportional to $u \delta N$. This can act as a driving force for the relative phase mode, as $\delta N \approx \hbar \chi_c \partial_t \phi/2$ for a small $u$. The order of magnitude estimate for these resonance frequencies are $\sim 0.1 T_c \sim 10$GHz.
 
We note that the layered structure of Sr$_2$RuO$_4$ will make these resonances easier to detect. Note that since we expect the $d$-vectors to point at the $c$-axis in the equilibrium, the Zeeman field will be applied along the $ab$ plane. Given the long penetration length ($\sim 152$nm) for this in-plane field, we do not expect the Meissner screening to be significant. 
Also, because the lower critical field is very small, we can actually induce a nearly uniform magnetic field.
  
The interorbital collective modes can also be detected is through relaxation processes. 
For instance, the phonon modes involving displacement of the next-nearest neighbor Ru atoms will modulate the interorbital coupling and hence can decay into the interorbital collective modes. Another is the NQR relaxation of the Ru atoms due to the relative phase mode, which 
gives rise to oscillating electric quadrupole moments on the Ru atoms as it involves oscillation of Cooper pair numbers between the Ru $d_{xz}$ and the Ru $d_{yz}$ orbitals. These electric quadrupole moments will relax the Ru nuclear quadrupole moments through the Ru atomic spin-orbit coupling. 
 
\section{Relation to collective modes of the 2D model}
\label{App:2Dmodel}
 
We show here how our collective modes are related to the collective modes studied for the 2D model. For every branch of collective modes in the 2D model, we can find its symmetry equivalent in the quasi-1D model. However, there are energy degeneracies in the 2D model which we expect would be broken in the quasi-1D model. We will show how this gives rise to the possibility of having soft collective modes in the quasi-1D model.
 
By generalizing the results from 3He-A \cite{WOLFLE1976}, one can see that there should be 12 branches of collective modes in a 2D chiral $p$-wave superconductors \cite{Tewordt1999, Higashitani2000}. Six of them involve no fluctuation of the orbital degrees of freedom, while the other six involves the relative fluctuation of the $p_x$ and $p_y$ pairings (termed `clapping' modes).
 
The spectra of the six branches involving no orbital fluctuation remain essentially the same regardless of whether we take the quasi-1D or the 2D model. They consist of the overall phase modes, the overall amplitude (Higgs) modes, the two branches of total spin modes, and also the two branches of condensate polarization modes. Regardless of the model, only the total spin modes can be soft; the overall amplitude and polarization modes have gaps equal to the pair-breaking frequency $2\Delta_0$ while the overall phase modes are gapped by the plasmon frequency.
 
On the other hand, the spectra of the $p_x$-$p_y$ relative modes are strongly affected by how close the system is to being rotationally invariant. There consist of six branches - the relative phase modes, the relative amplitude modes, the two branches of the relative spin modes, and the two branches the spin relative amplitude mode (the relative amplitude oscillation out of phase by $\pi$ between the spin up-up and down-down pairs). When the system is rotationally invariant, the relative phase modes and the relative amplitude modes are related by $\pi/4$ rotation around the $c$-axis due to the combined $U(1)$ symmetry of the overall phase and the orbital rotation. Therefore, up to the spin-orbit coupling, all six branches have the same gap, which is calculated to be $\sqrt{2}\Delta_0$ \cite{Higashitani2000, Kee2000}. However, when the rotational invariance is broken, this degeneracy between the relative phase and the relative amplitude mode is completely broken. Since the $p_x$ and the $p_y$ pairings are nearly decoupled in the quasi-1D model, the relative phase mode is nearly Goldstone while the relative phase mode have a gap that is almost same as the overall amplitude mode. Due to the same reason, the relative spin modes are nearly Goldstone, while the spin relative amplitude modes have a gap close to $2\Delta_0$.

\end{document}